\documentclass[prb,twocolumn,showpacs,tightenlines,10pt]{revtex4}
\usepackage{amssymb}
\usepackage{graphics}
\usepackage{epsfig}

\begin{document}

\title{Effects of La substitution on superconducting state of CeCoIn$_{5}$}
\author{C. Petrovic, S. L. Bud'ko, V. G. Kogan and P.C. Canfield}
\affiliation{Ames Laboratory and Department of Physics and Astronomy, Iowa State
University, Ames, Iowa 50011}
\date{\today}

\begin{abstract}
We report effects of La substitution on superconducting state of heavy
fermion superconductor CeCoIn$_{5}$, as seen in transport and magnetization
measurements. As opposed to the case of conventional superconductors, pair
breaking by nonmagnetic La results in depression of T$_{c}$ and indicates
strong gap anisotropy. Upper critical field H$_{c2}$ values decrease with
increased La concentration, but the critical field anisotropy, $\gamma
=H_{c2}^{a}/H_{c2}^{c}$, does not change in the Ce$_{1-x}$La$_{x}$CoIn$_{5}$
($x$=0-0.15). The electronic system is in the clean limit for all values of $%
x$.
\end{abstract}

\maketitle

\section{\label{Introduction}Introduction}

Heavy fermion superconductors have been extending for more than two decades
the affluence of physical phenomena associated with Cooper pair formation.%
\cite{FiskProc} The competition between magnetism and superconductivity for
the same Fermi surface of heavily renormalized carriers resulted in
observations of unconventional superconductivity\cite{BrulsUPt3}$^{,}$\cite%
{ShivaramUPt3} and raised speculations that spin pairing might be mediated
by magnetic interaction.\cite{PRB34Miyake6554} Research in the field was
associated with difficulties in sample preparation, sample to sample
variation, experimental conditions and ultimately, in the number of examples
where relevant physical phenomena can be observed in a clean form. The
recently discovered CeMIn$_{5}$ family (M=Ir, Rh, Co) of heavy fermion
superconductors encapsulates many aspects of important physics in this class
of materials. CeRhIn$_{5}$\cite{Rh} superconducts under applied pressures
above 17 kbar with T$_{c}\ $around 2 K whereas CeIrIn$_{5}$\cite{Ir} and
CeCoIn$_{5}$\cite{Co} are ambient pressure superconductors. CeCoIn$_{5}$
offers clean example of ambient pressure heavy fermion superconductivity
with a remarkably high T$_{c}$=2.3 K. The intriguing properties of CeCoIn$%
_{5}$ led to speculation that it may exhibit d-wave superconductivity,\cite%
{Kohori}$^{,}$\cite{NickCo}$^{,}$\cite{Izawa} and
Fulde-Ferrel-Larkin-Ovchinnikov state in high magnetic fields.\cite{Murphy}
In order to have more insight into the nature of CeCoIn$_{5}$ we perturbed
its superconducting state by substituting La onto the Ce site. For the
purpose of comparing influences of magnetic and nonmagnetic pair breaking on
T$_{c}$ suppression, we also substituted 5\% of Nd on Ce site. We find that
the anisotropy in the upper critical field does not change in the whole
concentration range and that the decrease of T$_{c}$ with increased La
doping cannot be explained solely with pressure effects due to unit cell
expansion. In addition, our results present an evidence for an anisotropic
order parameter in CeCoIn$_{5}$.

\section{\label{Experiment}EXPERIMENT}

Single crystals of Ce$_{1-x}$La$_{x}$CoIn$_{5}$ were grown by the self flux
method in a manner previously described.\cite{Co} Crystals grew as thin
plates with the c axis perpendicular to the plate. Removal of In from the
surface was performed by etching in concentrated HCl for several hours
followed by thorough rinsing in ethanol. All samples obtained with this
process showed no signs of In contamination. Powder X-ray patterns showed
that samples crystallized in HoCoGa$_{5}$ structure without any additional
peaks introduced by La alloying. In addition, magnetization measurements
provided a more sensitive test of possible presence of magnetically ordered
second phases. Both as grown and etched samples showed no sign of
antiferromagnetic transition of CeIn$_{3}$. Electrical contacts were made
with Epotek-H20E silver epoxy. In-plane resistivity was measured in Quantum
Design MPMS and PPMS\ measurement systems from 0.35 to 300\ K\ and in fields
up to 90 kOe applied parallel and perpendicular to the c-axis. There is
uncertainty in nominal resistivity values associated with sample geometry
and uneven surfaces of etched samples. We measured several samples for each
concentration in order to reduce measurement error which allowed us to
estimate uncertainties in nominal values as well. The dimensions of the
samples were measured by high precision optical microscope with 10$\mu $m
resolution and average values are presented. Randomly chosen samples within
each batch had no difference in their $R(T)$ curves. Magnetization
measurements were performed in MPMS-7 Quantum Design magnetometer in the
magnetic field of 10kG, applied parallel and perpendicular to c axis.

\section{\label{results}RESULTS}

The results of powder X-ray diffraction measurement taken at room
temperature are summarized in Table 1 and shown in Fig. 1, together with the
unit cell volume of LaCoIn$_{5}$. As expected, La doped samples have larger
unit cell volume. The volume increase in the concentration range $x=0-0.175$
is consistent with expansion of the unit cell as La substitutes Ce in
accordance with Vegard's law.

\begin{figure}
\epsfig{file=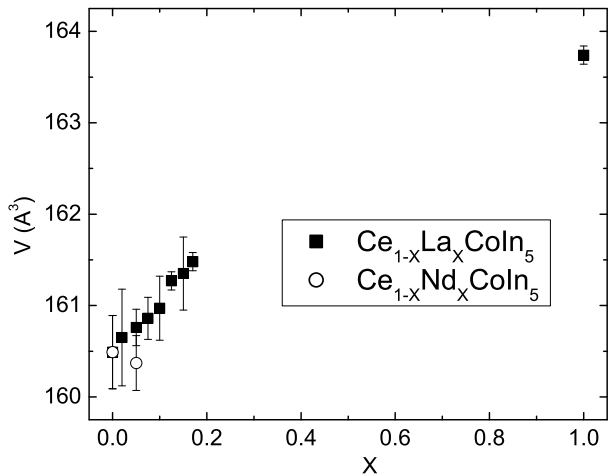,width=0.45\textwidth}
\caption{Unit cell volume of Ce$_{1-x}$La$_{x}$CoIn$_{5}$ (x=0-0.175, 1) shown
toghether with unit cell volume of Ce$_{0.95}$Nd$_{0.05}$CoIn$_{5}$}
\end{figure}

Fig. 2 shows the magnetic susceptibility for Ce$_{0.95}$Nd$_{0.05}$CoIn$_{5}$%
, Ce$_{0.85}$La$_{0.15}$CoIn$_{5}$, and CeCoIn$_{5}$, taken in the applied
field of 10kOe. In the whole temperature range above T$_{c}$, the
substitution of magnetic Ce$^{3+}$ by nonmagnetic La$^{3+}$ reduces
susceptibility values in La doped sample when compared with undoped CeCoIn$%
_{5}$. Comparison of high temperature moments through Curie-Weiss analysis
of the polycrystalline susceptibility average at high temperatures shows
that approximately 14\% Ce ions were substituted with La. No quantitative
difference from undoped CeCoIn$_{5}$ was detected in high temperature
susceptibility of 5\% Nd doped sample. Low temperature magnetic
susceptibility of Ce$_{0.85}$La$_{0.15}$CoIn$_{5}$ does not reveal any
difference in Curie tail from pure material, thus ruling out Kondo-hole
interpretation of La dilution (Fig. 2 inset).\cite{PRB53Lawrence12559} We
also see broadening of the plateau-like feature in $\chi _{c}$ in Ce$_{0.85}$%
La$_{0.15}$CoIn$_{5}$ ascribed\cite{NickCo} to thermal depopulation of Ce 4f
levels. On the other hand, Nd impurities contribute to pronounced Curie tail
at low temperatures. Subtraction of magnetic susceptibility of CeCoIn$_{5}$
from Ce$_{0.95}$Nd$_{0.05}$CoIn$_{5}$ in the normal state below 10K is
consistent with approximately 8\% of Nd$^{3+}$ paramagnetic moment, result
close to nominal stochiometric value and within rough approximation of our
analysis.

\begin{figure}
\epsfig{file=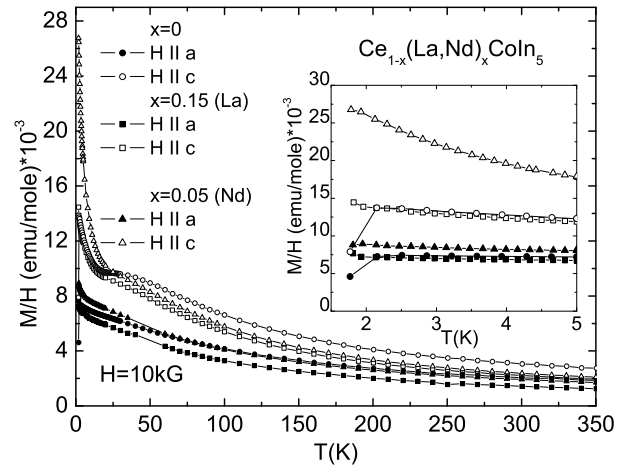,width=0.45\textwidth}
\caption{Magnetic
susceptibility of Ce$_{0.85}$La$_{0.15}$CoIn$_{5}$, Ce$_{0.95}$Nd$_{0.05}$%
CoIn$_{5}$ and CeCoIn$_{5}$. Low temperature susceptibility (inset) shows
pronounced Curie tail with 5\% of Nd substitution but no difference for 15\%
La substitution.}
\end{figure}
 
Temperature dependent electrical resistivities normalized to their
value at 300 K for Ce$_{1-x}$La$_{x}$CoIn$_{5}$ and Ce$_{0.95}$Nd$_{0.05}$%
CoIn$_{5}$ are presented in Fig. 3a. There are several key features to
notice. Resistivities of all samples are weakly temperature dependant at
high temperatures, and they pass through a maximum as temperature is
decreased. This behavior is traditionally interpreted as a crossover from
incoherent Kondo scattering to coherent Bloch states of heavy electrons in
the Kondo lattice. In the case of CeCoIn$_{5}$ this drop, at least
partially, could be attributed to depopulation of crystalline electric field
levels. We observe decrease of $T_{max}$ for higher La concentrations (Fig.
3a inset). At low temperatures, there is a clear suppression of T$_{c}$ as
more Ce ions are replaced by La (Fig. 3 inset). The increase of the\ normal
state residual resistivity $\rho _{0\text{ }}$is probably due to disorder
which contributes to increased conduction electron scattering. On the other
hand, the resistive transition width sharpens with La alloying. It is
interesting to note that Ce$_{1-x}$La$_{x}$CoIn$_{5}$ is not in the well
defined Fermi liquid regime above T$_{c}$: the $\rho (T)$ curves above T$%
_{c} $ do not show signs of $T^{2}$ dependence, as it has been reported for
CeCu$_{2}$Si$_{2}$.\cite{JLTP118Sheikin113} Depression of T$_{c}$ in CeCoIn$%
_{5}$ seems to scale with $\rho _{0\text{ }}$values for both magnetic and
nonmagnetic dopants, as seen by comparison of the $\rho (T)$ data of Ce$%
_{0.95}$Nd$_{0.05}$CoIn$_{5}$ and Ce$_{0.98}$La$_{0.02}$CoIn$_{5}$.

\begin{figure}
\epsfig{file=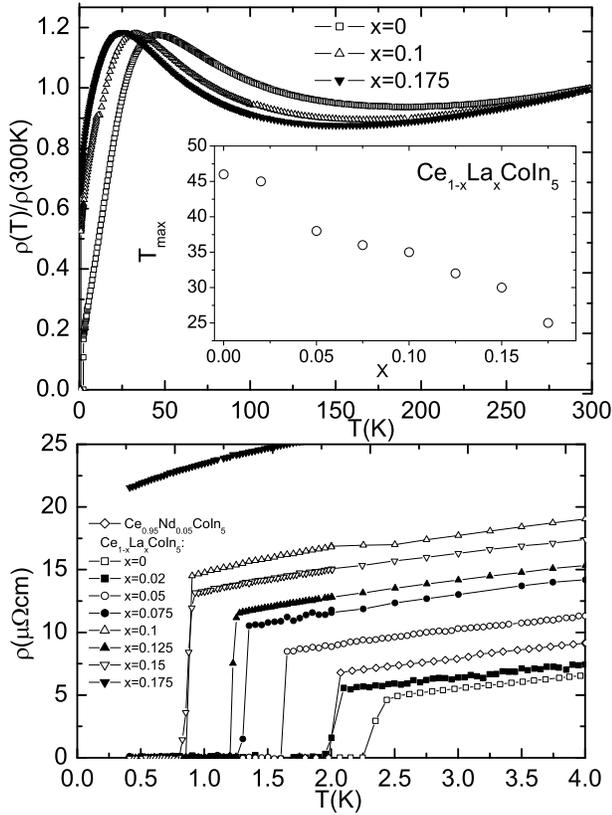,width=0.45\textwidth}
\caption{(a) Electrical resistivity $\protect%
\rho $ normalized to its value at 300 K vs. temperature for Ce$_{1-x}$La$%
_{x} $CoIn$_{5}$ for x = 0, 0.1 and 0.175. T$_{max}$ is shifted to lower
temperatures with increased La substitution (inset)\ (b) Low temperature
resistivity shows depression of T$_{c}$ and increase in $\protect\rho _{0}.$}
\end{figure}

Fig. 4 shows the anisotropic upper critical field for Ce$_{1-x}$La$_{x}$CoIn$%
_{5}$, normalized to transition temperature in zero field for each value of $%
x$ (values for $x=0$ were taken from previous report).\cite%
{JPSJ70Muramatsu3362} The H$_{c2}$ data were determined as a midpoint
between onset and zero in resistivity from $\rho (T)$ curves at constant
field and $\rho (H)$ curves at constant temperature. Adding La impurities
results in depression of H$_{C2}$, however, anisotropy $\gamma
=H_{c2}^{a}/H_{c2}^{c}$ remains at the same value of $\gamma \approx $ 2
(inset in Fig. 4). Uncertainty in our estimate of $\gamma $ decreases for
higher field data, away from H=0 transition (T/T$_{c}\approx $1).

\begin{figure}
\epsfig{file=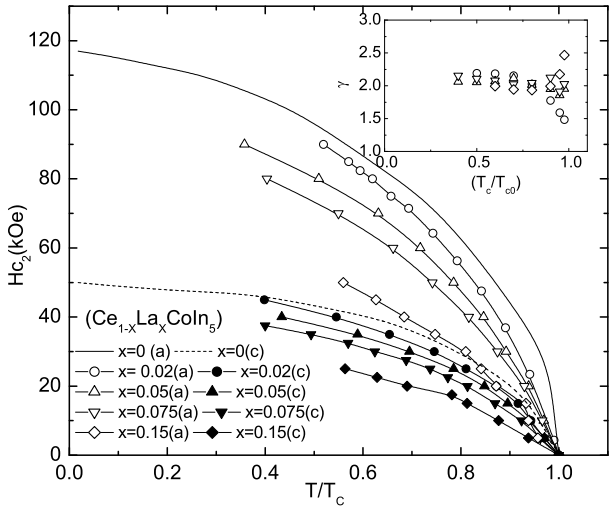,width=0.45\textwidth}
\caption{Anisotropy in the upper critical field
H$_{c2}$ for Ce$_{1-x}$La$_{x}$CoIn$_{5}$ (x = 0-0.15).\ Inset shows value
of $\protect\gamma $ = H$_{c2}^{a}/$H$_{c2}^{c}$ vs.T$_{c}$/T$_{c}$(H=0) for
varous La concentrations: x=0.02 (circles), x=0.05 (up triangles), x=0.075
(down triangles), x=0.15 (diamonds).}
\end{figure}

Assuming that Fermi surface properties of doped material do not change
substantially in the dilute La limit,\cite{PRB43Jee2656} it is reasonable to
assume inverse proportionality between $\rho $ and $l$, and therefore values
of $l_{0}$ could be estimated from $\rho _{0}$ for the whole doping series ($%
l_{0}=\frac{A}{\rho _{0}}$) using the the value of constant A from reported $%
l_{0\text{ }}$and $\rho _{0\text{ }}$values for pure material.\cite%
{PRL86Movshovich5152} We obtain $l_{0\text{ }}$ $\approx 540\mathring{A}$
for CeCoIn$_{5}$ without La impurities. Fig. 5 shows the ratio of mean free
path l$_{0}$ to in-plane superconducting coherence length $\xi $ $(\xi
^{2}(T)=\Phi _{0}/2\pi H_{c2}(T))$ for Ce$_{1-x}$La$_{x}$CoIn$_{5}$ obtained
at T=T$_{c}$/2. In the whole doping range electronic system is in the clean
limit which could explain nearly constant value of $\gamma
=H_{c2}^{a}/H_{c2}^{c}$.

\begin{figure}
\epsfig{file=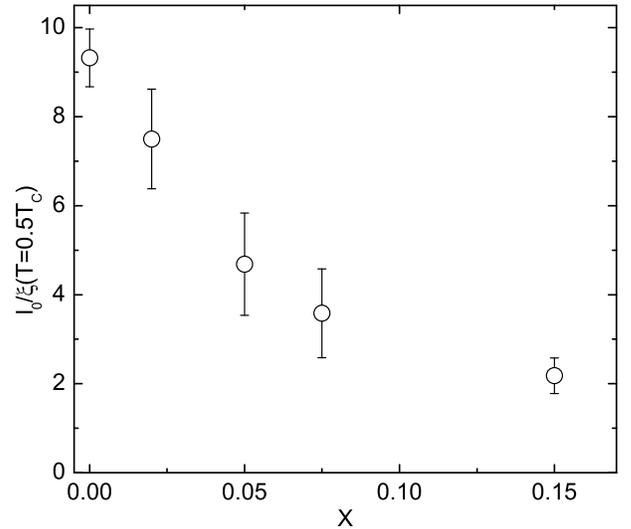,width=0.45\textwidth}
\caption{Ratio
of mean free path (l) to coherence length ($\protect\xi $) for Ce$_{1-x}$La$%
_{x}$CoIn$_{5}$. Electronic system is in the clean limit already at T=T$_{c}$%
/2 for La concentrations x = 0-0.15}
\end{figure}

A comparison of the effects of La substitution on T$_{c}$ in CeCoIn$_{5}$ and
CeCu$_{2.2}$Si$_{2}$ is shown on Fig. 6.\cite{JMMM76Ahlheim520} Doping
results in depression of T$_{c}$ in both cases but CeCoIn$_{5}$ is more
robust to pair breaking arising from La impurities. The initial rate of T$%
_{c}$ suppression is smaller than the rate seen in CeCu$_{2.2}$Si$_{2}$:
[(0.056T$_{c}$)/(1\% of La substitution) in CeCoIn$_{5}$ vs (0.085T$_{c}$%
)/(1\% of La substitution in CeCu$_{2.2}$Si$_{2}$)]. La doping in CeCoIn$%
_{5} $ is associated with only modest increase in nominal residual
resistivity values $\rho _{0}$, shown in Fig. 6 inset. The $\rho _{0}$
values for $x=0$ ($\sim $5$\mu \Omega $cm) in our experiment are in between
values reported previously in literature (3.1$\mu \Omega $cm\cite%
{PRL86Movshovich5152} and $\sim $7$\mu \Omega $cm\cite{JPCM13NicklasL905}).%

\begin{figure}
\epsfig{file=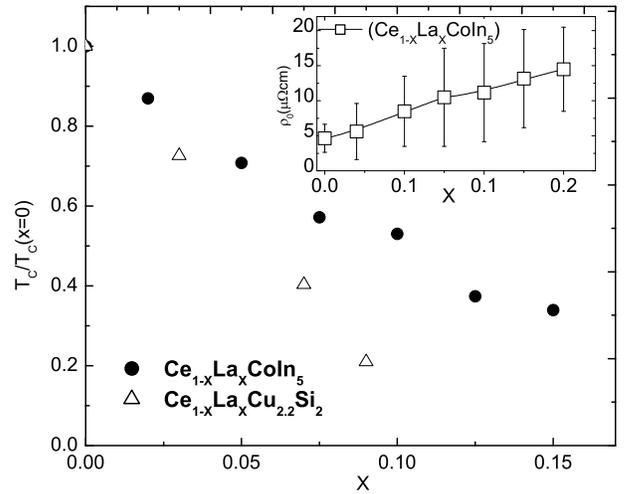,width=0.45\textwidth}
\caption{Comparison of La doping on T$%
_{c}$ of CeCoIn$_{5}$ (this work) and CeCu$_{2.2}$Si$_{2}$ (ref. 18). Inset
shows increase in $\protect\rho_{0}$ of Ce$_{1-x}$La$_{x}$CoIn$_{5}$ caused
by La substitution.}
\end{figure}

\section{\label{discussion}DISCUSSION AND CONCLUSIONS}

The slope of H$_{c2}$ vs T curve at T$_{c}$ can be used to estimate zero
temperature orbital critical field $H_{c2o}(0)$ using the weak - coupling
formula for conventional superconductors in Werthamer-Helfand-Hohenberg
model (WHH): $H_{c2o}(0)\approx 0.7(H_{c2}^{\prime })T_{c}$.\cite%
{PRWerthamer295} Table 1 shows estimates of $H_{c2}^{\prime }$ near T$_{c}$
for doped samples, together with previously reported value for $x=0$ for
both crystalline directions.\cite{JPSJ70Ikeda2248} All investigated samples
have high initial slopes, as expected in the case of heavy fermion
superconductors.\cite{PRB19Orlando4545}$^{,}$\cite{PRL49Rauchschwalbe1448}
Values of $H_{c2o}(0)$ decrease with introduction of La impurities (Table
1). The paramagnetic limiting field $H_{p}(0)=\Delta _{0}/\mu _{B}\sqrt{2%
\text{ }}$(where $\Delta _{0}$ is energy gap at T=0 and $\mu _{B}$ is Bohr
magneton) for pure CeCoIn$_{5}$ (T$_{c}$=2.3K) is well below\ the orbital
critical field $H_{c2o}(0)$ for either s-wave ($\Delta _{0}=3.52k_{B}T_{c}$),%
\cite{PRLClogston261} or d-wave pairing state ($\Delta _{0}=2.14k_{B}T_{c}$),%
\cite{PRB53Graf15147} and our results indicate that this unusual situation
is valid for investigated La doping range. We note that experimental values
of upper critical field for Ce$_{1-x}$La$_{x}$CoIn$_{5}$ (x=0-0.15) samples
are most likely below the values obtained by applying WHH model (Table 1),
probably due to polarization of magnetic sublattice due to enhanced internal
field along both crystalline axis.

\begin{table*}[tbp]
\caption{Properties of Ce$_{1-x}$La$_{x}$CoIn$_{5}$ doping series:\ T$_{c}$,
lattice parameters, unit cell volumes, H$_{c2}$'(T), calculated H$_{c2o}(0)$
from WHH\ model and approximate chemical pressure P$_{chemical}$ due to La
alloying. Final row: properties of Ce$_{0.95}$Nd$_{0.05}$CoIn$_{5}$.}%
\begin{tabular}{|l|l|l|l|l|l|l|l|}
\hline
$x$ & $T_{c}(K)$ & a($\mathring{A}$)($\pm 0.007\mathring{A})$ & c($\mathring{%
A}$)($\pm 0.007\mathring{A})$ & V($\mathring{A}$)$^{3}$ & $-\frac{dH_{c2}}{dT%
}(kOe/K)$ & $H_{c2o}(0)(kOe)$ & $P_{chemical}(kbar)$ \\ \hline
$0$ & 2.3 & 4.613 & 7.542 & 160.49$\pm 0.4$ & 240$(a),$110$\pm 6(c)$ & 
370(a), 170(c) & 0 \\ \hline
$0.02$ & 2.0 & 4.613 & 7.551 & 160.65$\pm 0.53$ & 170$\pm 23(a),$86$\pm 3(c)$
& 235(a), 119(c) & 0.6 \\ \hline
$0.05$ & 1.68 & 4.614 & 7.551 & 160.76$\pm 0.2$ & 190$\pm 19(a),$95$\pm 7(c)$
& 214(a), 107(c) & 1.1 \\ \hline
$0.075$ & 1.31 & 4.615 & 7.551 & 160.86$\pm 0.23$ & 207$\pm 27(a),$98$\pm
2(c)$ & 188(a), 89(c) & 1.5 \\ \hline
$0.1$ & 1.22 & 4.615 & 7.557 & 160.97$\pm 0.35$ &  &  & 2 \\ \hline
$0.125$ & 0.86 & 4.623 & 7.546 & 161.27$\pm 0.1$ &  &  & 3.1 \\ \hline
$0.15$ & 0.78 & 4.619 & 7.563 & 161.35$\pm 0.4$ & 236$\pm 27(a),$103$\pm
2(c) $ & 127(a), 55(c) & 3.5 \\ \hline
$0.175$ & - & 4.619 & 7.567 & 161.48$\pm 0.1$ &  &  &  \\ \hline
$1.0$ & - & 4.638 & 7.612 & 163.74$\pm 0.1$ &  &  &  \\ \hline
$0.05(Nd)$ & 2.0 & 4.601 & 7.546 & 160.37$\pm 0.3$ &  &  & -0.5 \\ \hline
\end{tabular}%
\end{table*}
It has recently been reported that T$_{c}$ in CeCoIn$_{5}$ increases under
applied pressure.\cite{JPCM13NicklasL905} Negative chemical pressure should
cause some decrease in T$_{c}$. In the lack of better approximation, we take
bulk modulus of CeCoIn$_{5}$ to be the same as the one for CeIn$_{3}$
(650kbar),\cite{ModulusCeIn3} and we calculate approximate chemical pressure
(P$_{chemical}$) for each La concentration using $\frac{V_{0}\partial P}{%
\partial V}\approx $650kbar. The results are shown in Table 1. Depression of
T$_{c}$ occurs at a rate $\frac{dT_{c}}{dP}\approx 0.43$ K/kbar - a slope
that is an order of magnitude larger than reported increase of T$_{c}$ under
hydrostatic pressure. An order of magnitude difference from pure pressure
effect on T$_{c}$ is likely to exceed error in estimation of bulk modulus,
and therefore points to the conclusion that the pair breaking mechanisms
which enter through disorder due to La alloying and increased scattering of
Cooper pairs are dominant in CeCoIn$_{5}$. In contrast to conventional
superconductors where nonmagnetic impurities have small effect on T$_{c}$,
Cooper pairs formed in CeCoIn$_{5}$ are rather sensitive to La doping: 2\%
of La depresses T$_{c}$ to the same value as $\sim $5\% of Nd.

The T$_{c}$ suppression induced by the nonmagnetic La substitution in Ce$%
_{1-x}$La$_{x}$CoIn$_{5}$ is reminiscent of the pair breaking effect by
magnetic impurities.\cite{JETP12Abrikosov1243} Although various factors may
suppress T$_{c}$ (an anisotropic scattering, for example),\cite{Joerg} we
focus here on the scenario of CeCoIn$_{5}$ having an anisotropic gap $\Delta
(\vec{k}_{F})$ on the Fermi surface. This scenario is quite likely given the
unconventional nature in many heavy-fermion materials.

It is known\cite{JETP4Hohenberg1208} that if $\Delta $ depends on the
position at the Fermi surface, the critical temperature is suppressed by
nonmagnetic scattering according to:

\begin{equation}
ln\frac{T_{c0}}{T_{c}}=\alpha \left[ \psi (\frac{1+\mu }{2})-\psi (\frac{1}{2%
})\right] ,\mu =\frac{\hbar }{2\pi T_{c}}
\end{equation}

Here T$_{c0}$ is the critical temperature of the material in the absence of
all scattering, $\tau $ is the scattering time by nonmagnetic impurities,
and $\alpha =1-\left\langle \Delta \right\rangle ^{2}/\left\langle \Delta
^{2}\right\rangle $ characterizes the gap anisotropy, $\left\langle
...\right\rangle $ stands for averaging over Fermi surface, and $\psi $ is
the digamma function. For a weak gap anisotropy, this result is due to
Hohenberg,\cite{JETP4Hohenberg1208} see also later publications.\cite%
{PR131Markowitz563}$^{,}$\cite{JETP63Posazhennikova347} It can be shown that
in fact Eq. (1) holds for an arbitrary gap anisotropy.\cite{Koganunpublished}
For isotropic $\Delta $, $\alpha =0$, and we come to Anderson's theorem: T$%
_{c}$=T$_{c0}$. For pure d-wave order parameter, $\left\langle \Delta
\right\rangle =0$, and Eq. (1) describes the d-pair breaking by nonmagnetic
scattering (which differs from the Abrikosov-Gor'kov result only by the
factor of 2 in the definition of the parameter $\mu _{m}$=$\hbar /\pi
T_{c}\tau _{m}$).

To analyze the $T_{c}(x)$ data shown in Fig. 6, one has to relate $x$ to the
scattering time $\tau $, a nontrivial connection. We avoid this difficulty
by assuming that the residual resistivity $\rho _{0}$ is proportional to 1/$%
\tau $. Further, we exclude parameter T$_{c0}$ from Eq. (1) by writing it
for two values of $x$ and subtracting the results:%
\begin{equation}
ln\frac{T_{_{2}}}{T_{_{1}}}=\alpha \lbrack \psi (\frac{1+\mu _{_{1}}}{2}%
)-\psi (\frac{1+\mu _{2}}{2})],\mu _{_{1,2}}=\beta \frac{\rho _{_{1,2}}}{%
T_{_{1,2}}}
\end{equation}%
where $T_{1,2}=T_{c}(x_{1,2})$ and $\beta $ is a constant to be determined.
Writing this equation for two different pairs $x_{1,2}$ one can determine
the unknown $\alpha $ and $\beta $. This procedure yields values scattered
around $\alpha =0.5$ and $\beta =0.2K/\mu \Omega cm$.

Hence, we find $\alpha =\left\langle \Delta \right\rangle ^{2}/\left\langle
\Delta ^{2}\right\rangle \approx 0.5$ which implies a strongly anisotropic
gap. Knowing the value of $\beta $ we can estimate the scattering time using
measured resistivities; for $x=0$ we obtain $\tau =\hbar /2\pi k_{B}\beta
\rho \approx 1.3\times 10^{-12}s.$ With the electronic specific heat
coefficient\cite{PRL86Movshovich5152} $\gamma =290mJ/moleK^{2}$ we roughly
estimate the Fermi velocity $v_{F}=\frac{\pi k_{B}}{e\sqrt{\gamma \tau \rho
_{0}}}\approx 2\times 10^{6}$cm/s. This would correspond to the mean-free
path $l\approx 260\mathring{A}$, a value smaller than expected but within
factor of two of our determination of mean free path which is reasonable
given the assumptions of average Fermi velocity and isotropic scattering.

In summary, diamagnetic pair breaking effect in CeCoIn$_{5}$ is consistent
with picture of strongly anisotropic order parameter. Anisotropy in the
upper critical field $\gamma =H_{c2}^{a}/H_{c2}^{c}$ does not change for x =
(0-0.15) in Ce$_{1-x}$La$_{x}$CoIn$_{5}$, indicating electronic system in
the clean limit.

We thank Joerg Schmalian for useful discussions and Hal Sailsbury for help
with optical microscope. This work was carried out at Ames Laboratory, which
is operated for the U.S. Department of Energy by Iowa State University under
Contract No. W-7405-82. This work was supported by the Director for Energy
Research, Office of Basic Energy Sciences of the U.S. Department of Energy.

\end{document}